\documentclass[12pt]{iopart}
\usepackage{graphicx}
\usepackage{dcolumn}
\usepackage{bm}

\begin{document} 

\title[Burnt-Bridge Model]{Dynamic Properties of Molecular Motors in Burnt-Bridge Models}
\author{Maxim N. Artyomov\ddag, Alexander Yu. Morozov\dag, Ekaterina Pronina\dag,  and Anatoly B. Kolomeisky\dag}
\address{\dag Department of Chemistry, Rice University, Houston, TX 77005, USA}
\address{\ddag Department of Chemistry, Massachusetts Institute of Technology, Cambridge, MA 02139 USA}

\begin{abstract}
Dynamic properties of molecular motors that fuel their motion by actively interacting with underlying molecular tracks are studied theoretically via  discrete-state stochastic ``burnt-bridge'' models. The transport of the particles is viewed as an effective  diffusion along one-dimensional lattices with periodically distributed weak links. When an unbiased random walker passes the weak link it can be destroyed (``burned'') with probability $p$, providing a bias in the motion of the molecular motor. A new theoretical approach that allows one to calculate exactly {\it all} dynamic properties of motor proteins, such as velocity and dispersion, at general conditions is presented.  It is found that dispersion is a decreasing  function of the concentration of bridges, while the dependence of dispersion on the burning probability is more complex. Our calculations also show a gap in dispersion  for very low concentrations of weak links  which indicates a dynamic phase transition between unbiased and biased diffusion regimes. Theoretical findings are supported by Monte Carlo computer simulations.

\end{abstract}

\ead{tolya@rice.edu}

\maketitle

\section{Introduction}

Recently much attention has been devoted to experimental and theoretical studies of motor proteins, that are active enzyme molecules that move along linear molecular tracks by consuming a chemical energy and transforming it into a mechanical work \cite{AR}. Motor proteins play important role in many biological processes \cite{lodish_book,howard_book,bray_book}. While conventional molecular motors are powered by the hydrolysis of adenosine triphosphate (ATP) or related compounds, it was shown recently  that a  protein collagenase utilizes a different mechanism for the transport along the collagen fibrils \cite{saffarian04,saffarian06}.  It fuels its motion by a  proteolysis, i.e., the collagenase catalyzes the cutting of a protein filament  at some specific positions, and the enzyme molecule is always found on one side of the proteolysis site. Since the collagenase molecule cannot cross the broken site once the cleavage of the filament occurred, it leads to the biased diffusion of the motor protein along the fibril.

It was suggested that the unusual dynamics of collagenase  can be described by the so-called ``burnt-bridge model'' (BBM) \cite{saffarian04,saffarian06,mai01,antal05,morozov07}.  In BBM  the motor protein molecule is viewed as an unbiased random walker that hops along the one-dimensional lattice composed of strong and weak links. Strong links are not affected when crossed by the walker, whereas the weak links (``bridges'') might be destroyed (``burnt'') with a probability $0<p\le 1$ when the walker passes over them; and the burnt bridges cannot be crossed again. However, theoretical studies of dynamic properties of motor proteins in BBM are still very limited. Limiting cases of  very low burning probabilities and $p \rightarrow 1$ case have been analyzed earlier by using spatial continuum approximation \cite{mai01}, which is valid only for very low concentrations of bridges. But experimental studies \cite{saffarian04,saffarian06} suggest that the density of proteolysis sites on collagen is significant. A different approach was utilized by Antal and Krapivsky \cite{antal05}, who investigated discrete-time dynamics of motor proteins.  They calculated the mean drift velocity of the particle $V$ for periodically distributed bridges  for the entire range of parameters, while the dispersion $D$ has been determined only for the special simple case of $p=1$ for both periodic and random bridge distributions. However, it was pointed out later \cite{morozov07} that the realistic description of motor protein dynamics requires a continuum-time description since motor proteins move following Poisson statistics.

Recently, we developed a theoretical approach to investigate continuous-time and discrete-time dynamics of collagen motor proteins in BBM \cite{morozov07} for the periodic distribution of weak links. First, it was shown that the special case of $p=1$ corresponds to the motion of the single particle on infinite periodic lattices \cite{derrida83}, for which all dynamic properties can be obtained exactly. For the general case of $p<1$ we developed a simple method by considering a reduced chemical kinetic scheme,  and some dynamic properties have been explicitly calculated. However, there is a deficiency  with this method. In this approach the analytical expressions for the dispersion and  correlation functions have not been obtained, although the simultaneous knowledge of the velocity and dispersion is crucial for understanding mechanisms of motor protein transport \cite{AR,morozov07}. The goal of the present work is to develop a comprehensive theoretical method that allows one to determine {\it all} dynamic properties of motor proteins in BBM for general sets of parameters.

\section{Model}

In our model the single motor protein is represented by a particle that jumps along an infinite one-dimensional lattice with transition rates in both directions assumed to be equal to one, as illustrated in Fig. 1. The size of the lattice spacing is also taken to be equal to one. The lattice consists of strong and weak bonds. Whereas the random walker has no effect on the strong links, it destroys the weak ones with the probability $p$ when crossing them. After the bridge is burnt, the particle is always assumed to be to the right of it. In addition, it is postulated that all bridges are intact at $t=0$. Therefore, the trapping of the walker between two burnt bridges cannot occur, the particle continuously moves to the right.

\begin{figure}[ht]
\centering
\includegraphics[scale=0.6,clip=true]{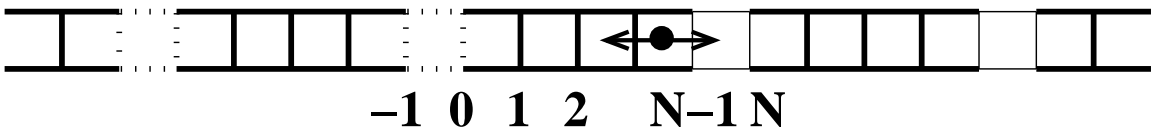}
\caption{Schematic picture for the burnt-bridge model of the transport of single particles. Thick solid lines describe lattice cells with strong links that cannot be burned. Thin solid lines correspond to the weak links that can be burned with the probability $p$ when the particle crosses them from  left to right. Already burned bridges are shown with dotted lines. The particle cam hop with equal rates one  step forward or backward   unless the link is already destroyed.}
\end{figure}

The dynamic properties of the molecular motors in  BBM strongly depend on the details of the bridge burning \cite{morozov07}. There are two possibilities of breaking weak links. In the first model, which we term a forward BBM, the bridge burns only when crossed from left to right, but it always stays intact if crossed in the opposite direction. In the second model, which describes a forward-backward BBM, the bridge is burnt when crossed in either direction. Both models are identical for the case of $p=1$ because of irreversible burning.  However, for the general case of $p<1$ the walker's dynamics in two scenarios, although  qualitatively similar, can show some   difference \cite{morozov07}. For simplicity, in this paper we will only analyze forward BBM, although our method can be easily extended for forward-backward BBM.

Two parameters specify the dynamics of the molecular motors in BBM: the burning probability $p$ and the concentration of bridges $c$. Another important factor that strongly influences the dynamic properties is the distribution of bridges \cite{antal05}. Two possible distributions of weak links, periodic and random, have been considered so far \cite{antal05}. In this paper we will investigate  the periodic bridge distribution, in which the bridges are located at a constant distance $N=1/c$ lattice spacings apart from each other. Thus weak links can be found at the lattice sites with the coordinates $kN$, where $k$ is integer (see Fig. 1). Periodic distribution of bridges is more realistic for the description of dynamics of collagenases \cite{saffarian04,saffarian06}.

\section{Computation of Dynamic Properties}

Our theoretical approach is a generalization of the method proposed by Derrida for explicit calculation of dynamic properties of the random walker on infinite periodic one-dimensional lattices \cite{derrida83}. It was shown earlier that the original version of Derrida's method can be directly applied to the special case of BBM with $p=1$ (with periodic bridge distribution) \cite{morozov07}, and it leads to analytical computations of all dynamic properties of the system in this special case. However, the direct application of the method cannot be done for the case of the bridge burning with the probability $p < 1$, since it is unclear how the transition rates depend on $p$. We propose a method that explicitly takes into account the presence of the weak links, the feature that is not accounted for in the application of the original Derrida's method. 

Let us define $P_{kN+i,y}(t)$ as the probability that at time $t$ the walker is at point $x=kN+i$ ($i=0,1,\cdots,N-1$) given that the right end of the last burnt bridge is at the point $yN$. We note that $y$ and $k \ge 0$ are some integers, and 
\begin{equation}\label{r0}
R_{kN+i}(t)=\sum \limits_{y=-\infty}^{+\infty}P_{(y+k)N+i,y}(t), \quad  (k \ge 0; \ i=0,1,\cdots,N-1),
\end{equation}
where $R_{kN+i}(t)$ is the probability to find the random walker $kN+i$ sites apart from the last burnt bridge at time $t$. It was found in Ref. \cite{morozov07} that  in the  stationary-state limit ($t \rightarrow \infty$) the explicit expressions for these functions are given by
\begin{equation}\label{rkn}
R_{kN+i}(t \rightarrow \infty)=R_{kN+i} = R_{0} x^k\left[1-\frac{i}{N}(1-x)\right],
\end{equation}
where 
\begin{equation}\label{rr0}
R_0 = \frac{2p}{\sqrt{p^2(N-1)^2 + 4pN}} = \frac{2pc}{\sqrt{p^2(1-c)^2 + 4pc}},
\end{equation}
and
\begin{equation}\label{x}
x = 1+\frac{1}{2}p(N-1)-\frac{1}{2}\sqrt{p^2(N-1)^2 + 4pN}.
\end{equation}
Following the  Derrida's method \cite{derrida83},  we introduce additional auxiliary functions $S_{kN+i}(t)$,
\begin{equation}\label{s0}
S_{kN+i}(t)=\sum\limits_{y=-\infty}^{+\infty}(yN+kN+i)P_{yN+kN+i,y}(t), \quad (k \ge 0 \ i=0,1,\cdots,N-1).
\end{equation}

The dynamics of the system is governed by a set of Master equations:
\begin{equation}\label{p1}
\frac{d P_{yN,y}(t)}{d t}= P_{yN+1,y}(t)-P_{yN,y}(t)+ p \sum_{y^{\prime}=-\infty}^{y-1} P_{yN-1,y^{\prime}}(t),
\end{equation}
for $k=i=0$, and 
\begin{equation}\label{p2}
\frac{d P_{yN+kN,y}(t)}{dt}=P_{yN+kN+1,y}(t)+(1-p)P_{yN+kN-1,y}(t)-2P_{yN+kN,y}(t),
\end{equation}
for $k \ge 1$ and $i=0$. Also,
\begin{equation}\label{p3}
\frac{d P_{yN+kN+i,y}(t)}{dt}=P_{yN+kN+i+1,y}(t)+P_{yN+kN+i-1,y}(t)-2P_{yN+kN+i,y}(t)
\end{equation}
for $k \ge 0$ and $i=1,\cdots,N-1$. It should be noted that substituting the  definition (\ref{r0}) into Eqs. (\ref{p1}) - (\ref{p3}) leads to the equations for $R_{kN+i}$ that have been already  obtained in  Ref. \cite{morozov07}, although through the different route,
\begin{equation} \label{r1}
\frac{d R_{0}(t)}{d t}=p \sum_{k=1}^{\infty} [R_{kN-1}(t)] +R_{1}(t)-R_{0}(t),
\end{equation}
\begin{equation}\label{r2}
\frac{d R_{kN}(t)}{dt}=(1-p)R_{kN-1}(t)+R_{kN+1}(t)-2R_{kN}(t) 
\end{equation}
for $k \ge 1$ and 
\begin{equation}\label{r3}
\frac{d R_{kN+i}(t)}{dt} = R_{kN+i-1}(t)+R_{kN+i+1}(t)-2 R_{kN+i}(t),
\end{equation}
for $k \ge 0$ and $ i=1,\cdots,N-1$.

In the similar way, the time evolution of the functions $S_{kN+i}(t)$ can be obtained  from Eqs. (\ref{s0})  and Eqs. (\ref{p1}) - (\ref{p3}). After some rearrangement it can be shown that
\begin{equation}\label{s1}
\frac{d S_{0}(t)}{d t}=S_1(t)-S_0(t)+p\sum\limits_{\alpha=1}^{\infty}S_{\alpha N-1}(t)+p\sum\limits_{\alpha=1}^{\infty}R_{\alpha N-1}(t)-R_1(t),
\end{equation}
\begin{equation}\label{s2}
\frac{d S_{kN}(t)}{d t}=S_{kN+1}(t)+(1-p)S_{kN-1}(t)-2S_{kN}(t)+(1-p)R_{kN-1}(t)-R_{kN+1}(t)
\end{equation}
for $k \ge 1$ and 
\begin{equation}\label{s3}
\frac{d S_{kN+i}(t)}{d t}=S_{kN+i+1}(t)+S_{kN+i-1}(t)-2S_{kN+i}(t)+R_{kN+i-1}(t)-R_{kN+i+1}(t)
\end{equation} 
for $k \ge 0$ and $i=1,\cdots,N-1$.

At large times the solutions of Eqs. (\ref{s1}) - (\ref{s3}) can be found assuming the following form,
\begin{equation}\label{sj}
S_j(t)=a_j t+T_{j},
\end{equation} 
with time-independent coefficients $a_j$ and $T_j$. Substituting Eq. (\ref{sj}) into  Eqs. (\ref{s1}) - (\ref{s3}) yields,
\begin{equation}\label{a1}
a_1-a_0+ p \sum\limits_{\alpha=1}^{\infty}a_{\alpha N-1}=0,
\end{equation}
\begin{equation}\label{a2}
a_{kN+1}+(1-p)a_{kN-1}-2a_{kN}=0  
\end{equation}
for $k \ge 1$ and
\begin{equation}\label{a3}
a_{kN+i+1}+a_{kN+i-1}-2a_{kN+i}=0 
\end{equation}
for $k \ge 0$ and $ i=1,\cdots,N-1$.

One can easily see  that Eqs. (\ref{a1}) - (\ref{a3}) are identical to the system of equations for the functions $R_j$ [Eqs. (\ref{r1}) - (\ref{r3})] in the stationary-state limit (with $dR_j /dt=0$). Therefore their solutions are the same up to the multiplicative constant, i.e.,
\begin{equation}\label{c}
a_{kN+i}=C R_{kN+i},
\end{equation}
where $R_{kN+i}$ is given by Eq. (\ref{rkn}). Because of the normalization condition  $\sum\limits_{k=0}^{\infty}\sum\limits_{i=0}^{N-1} R_{kN+i}=1$ \cite{morozov07}, it follows that $C=\sum\limits_{k=0}^{\infty}\sum\limits_{i=0}^{N-1} a_{kN+i}$. To find it explicitly, it is convenient to write the equations for $T_j$ which result from the substitution of Eq. (\ref{sj}) into Eqs. (\ref{s1}) - (\ref{s3}),
\begin{equation}\label{t1}
a_0=T_1-T_0+p\sum\limits_{\alpha=1}^{\infty}T_{\alpha N-1}+p\sum\limits_{\alpha=1}^{\infty}R_{\alpha N-1}-R_1,
\end{equation}
\begin{equation}\label{t2}
a_{kN}=T_{kN+1}+(1-p)T_{kN-1}-2T_{kN}+(1-p)R_{kN-1}-R_{kN+1},
\end{equation}
for $k \ge 1$ and 
\begin{equation}\label{t3}
a_{kN+i}=T_{kN+i+1}+T_{kN+i-1}-2T_{kN+i}+R_{kN+i-1}-R_{kN+i+1},
\end{equation} 
for $k \ge 0$ and $ i=1,\cdots,N-1$. Summing over  Eqs. (\ref{t1}) - (\ref{t3}) produces 
\begin{equation}\label{c1}
C=\sum\limits_{k=0}^{\infty}\sum\limits_{i=0}^{N-1} a_{kN+i}=R_0.
\end{equation}
Therefore, according to Eq. (\ref{c})
\begin{equation}\label{akn}
a_{kN+i}=R_0 R_{kN+i}
\end{equation} 
with $R_{kN+i}$ and $R_0$ given by Eqs. (\ref{rkn}) and  (\ref{rr0}).

Based on this analysis  it is now possible to obtain the expression for the velocity of the particle. The mean position of the random walker is the following,
\begin{eqnarray}\label{avx}
\langle x(t)\rangle=\sum\limits_{y=-\infty}^{+\infty}\sum\limits_{k=0}^{\infty}\sum\limits_{i=0}^{N-1} (yN+kN+i)P_{yN+kN+i,y}(t)= \nonumber\\ \sum\limits_{k=0}^{\infty}\sum\limits_{i=0}^{N-1}\left\{\sum\limits_{y=-\infty}^{+\infty} (yN+kN+i)P_{yN+kN+i,y}(t)\right\}=\sum\limits_{k=0}^{\infty}\sum\limits_{i=0}^{N-1} S_{kN+i}(t).
\end{eqnarray} 
Therefore the mean velocity is given by
\begin{equation}\label{v1}
V=\frac{d}{dt} \langle x(t)\rangle=\sum\limits_{k=0}^{\infty}\sum\limits_{i=0}^{N-1} \frac{d}{dt} S_{kN+i}(t).
\end{equation} 
In the large-time limit it can be shown from Eqs. (\ref{sj}) and  (\ref{c1}) that
\begin{equation}\label{vv}
V(c,p)=\sum\limits_{k=0}^{\infty}\sum\limits_{i=0}^{N-1} a_{kN+i}=R_0,
\end{equation} 
with $R_0$  explicitly given in Eq. (\ref{rr0}). Note that, as expected, this expression  reproduces the result for $V(c,p)$ already obtained in Ref. \cite{morozov07} via the reduced chemical kinetic scheme method.

In order to  calculate the diffusion coefficient $D(c,p)$, it is necessary to find functions $T_{kN+i}$ from Eqs. (\ref{t1}) - (\ref{t3}). To this end, it is convenient to rewrite Eqs. (\ref{t1}) - (\ref{t3}) by replacing $a_{kN+i}$ according to Eqs. (\ref{akn}) and (\ref{rkn}), and using the following relations,
\begin{equation}\label{}
p\sum\limits_{\alpha=1}^{\infty}R_{\alpha N-1}-R_1=\frac{2R_0}{N}(1-x)-R_0,
\end{equation} 
\begin{equation}\label{}
(1-p)R_{kN-1}-R_{kN+1}=R_{kN+i-1}-R_{kN+i+1}=\frac{2R_0}{N}(1-x)x^k,
\end{equation} 
that follow from Eq. (\ref{rkn}) and  Eqs. (\ref{r1}) - (\ref{r3}) in the stationary-state limit. As a result, Eqs. (\ref{t1}) - (\ref{t3}) are transformed into the following expressions,
\begin{equation}\label{tt1}
T_1-T_0+p\sum\limits_{\alpha=1}^{\infty}T_{\alpha N-1}+\Gamma-R_0=0,
\end{equation}
\begin{equation}\label{tt2}
T_{kN+1}+(1-p)T_{kN-1}-2T_{kN}+\Gamma x^k=0
\end{equation}
for $k \ge 1$ and 
\begin{equation}\label{tt3}
T_{kN+i+1}+T_{kN+i-1}-2T_{kN+i}+\Gamma x^k+\frac{i}{N}(1-x)R_0^2 x^k=0
\end{equation} 
for $k \ge 0$ and $ i=1,\cdots,N-1$. In Eqs. (\ref{tt1}) - (\ref{tt3}) we defined a new function $\Gamma$,
\begin{equation}\label{gamma}
\Gamma =\frac{2R_0}{N}(1-x)-R_0^2,
\end{equation} 
with $R_0$ and $x$ given by Eqs. (\ref{rr0}) and (\ref{x}), correspondingly.

We are looking  for the solution of the Eqs. (\ref{tt1}) - (\ref{tt3}) in the following form, 
\begin{equation}\label{tt0}
T_{kN+i}=(Ak+B)x^k\left[1-\frac{i}{N}(1-x)\right]-\frac{i(i+1)}{2}\Gamma x^k -\frac{i(i^2-1)}{6}\frac{(1-x)}{N}R_0^2 x^k -\Delta x^k.
\end{equation} 
There are three unknown variables in Eq. (\ref{tt0}): $A$, $B$ and $\Delta$. It is easy to see that regardless of their values, Eq. (\ref{tt0}) automatically solves Eq. (\ref{tt3}). The $B$-term solves the homogeneous system [Eqs. (\ref{tt1}) - (\ref{tt3}) without $\Gamma$- and $R_0$-terms] as follows from the comparison of Eqs. (\ref{tt1}) - (\ref{tt3}) with Eqs. (\ref{a1}) - (\ref{a3}) and Eq. (\ref{rkn}). As we show below, the value of $B$ does not affect the result for the dispersion  $D(c,p)$ and it may remain undetermined. Therefore, there are two unknowns, $A$ and $\Delta$, that need to be determined from Eqs. (\ref{tt1}) - (\ref{tt2}). Substituting Eq. (\ref{tt0}) into Eqs. (\ref{tt1})  and (\ref{tt2}) yields the system of equations for $A$ and $\Delta$:
\begin{eqnarray}\label{sys1}
\frac{A}{N}(1-p)[1+(N-1)x]+\Delta [-x+(1-p)]&=&-(1-p)\alpha, \\ \label{sys2}
A\frac{p}{N}[1+(N-1)x]\frac{x}{(1-x)^2}-\Delta \frac{p}{1-x}&=&\frac{p\alpha}{1-x}+R_0,
\end{eqnarray}
with 
\begin{equation}\label{alpha}
\alpha =\frac{N(N-1)}{2}\Gamma +\frac{(N-1)(N-2)}{6}(1-x)R_0^2,
\end{equation} 
while the parameters $\Gamma$, $R_0$, $x$ are given by Eqs. (\ref{gamma}), (\ref{rr0}), and (\ref{x}).

Solving the system  of Eqs. (\ref{sys1}) - (\ref{sys2}) results in the following expressions for $A$ and $\Delta$:
\begin{eqnarray}\label{A}
A &=&\frac{N(1-x)[\alpha px+R_0(1-x)(-1+p+x)]}{p[1+(N-1)x](-1+p+x^2)}, \\
\label{delta}
\Delta &=&\frac{(1-p)[\alpha p+R_0(1-x)^2]}{p(-1+p+x^2)}.
\end{eqnarray} 
Since $R_0$, $x$ (and hence $\Gamma$ and $\alpha$) are functions of $(c,p)$, Eqs. (\ref{A}) and (\ref{delta}) provide the expressions for $A(c,p)$ and $\Delta(c,p)$. Then substituting them into Eq. (\ref{tt0}) yields the formula for $T_{kN+i}(c,p)$.

It should be pointed out that Eq. (\ref{tt0}) is not the only possible form of $T_{kN+i}$ that solves the system of Eqs. (\ref{tt1}) - (\ref{tt3}). As an alternative, one can use, for example, $\tilde{T}_{kN+i}$ which is the same as (\ref{tt0}) except for the last term when $\Delta x^k$ in Eq. (\ref{tt0}) is replaced by $i\tilde{\Delta} x^k$. As a result, corresponding expressions for $\tilde{A}(c,p)$ and $\tilde{\Delta}(c,p)$ can be obtained. However, these expressions are much more cumbersome than those given by Eqs. (\ref{A}) and (\ref{delta}), and the final resulting equation for the diffusion coefficient $D(c,p)$, as can be shown, does not depend on the specific choice of the functions  $T_{kN+i}$.

In order to calculate the diffusion coefficient $D(c,p)$ additional auxiliary functions are introduced, following Derrida's method \cite{derrida83},
\begin{equation}\label{ukn}
U_{kN+i}(t)=\sum\limits_{y=-\infty}^{+\infty}(yN+kN+i)^2 P_{yN+kN+i,y}(t), \quad k \ge 0, \quad  i=0,1,\cdots,N-1.
\end{equation} 
The next step is to determine a set of equations that govern the time evolution of $U_j(t)$. Corresponding derivation is similar to that of the Eqs. (\ref{s1}) - (\ref{s3}) for  functions $S_{kN+i}$. Using the definitions of $U_{kN+i}$, $S_{kN+i}$ and $R_{kN+i}$, given by Eqs. (\ref{ukn}), (\ref{s0}) and (\ref{r0}) together with Eqs. (\ref{p1}) - (\ref{p3}), leads to the following expressions,
\begin{eqnarray}\label{u1}
\frac{d U_{0}(t)}{d t}=U_1(t)-U_0(t)-2S_1(t)+R_1(t)+p\sum\limits_{\alpha=1}^{\infty}U_{\alpha N-1}(t) \nonumber \\
+2p\sum\limits_{\alpha=1}^{\infty}S_{\alpha N-1}(t)+p\sum\limits_{\alpha=1}^{\infty}R_{\alpha N-1}(t),
\end{eqnarray}
\begin{eqnarray}\label{u2}
\frac{d U_{kN}(t)}{d t}=U_{kN+1}(t)+(1-p)U_{kN-1}(t)-2U_{kN}(t)-2S_{kN+1}(t) \nonumber \\
+2(1-p)S_{kN-1}(t)+R_{kN+1}(t)+(1-p)R_{kN-1}(t)
\end{eqnarray}
for $k \ge 1$, and 
\begin{eqnarray}\label{u3}
\frac{d U_{kN+i}(t)}{d t}=U_{kN+i+1}(t)+U_{kN+i-1}(t)-2U_{kN+i}(t)-2S_{kN+i+1}(t) \nonumber \\
+2S_{kN+i-1}(t)+R_{kN+i+1}(t)+R_{kN+i-1}(t)
\end{eqnarray} 
for $k \ge 0$ and $ i=1,\cdots,N-1$.

The diffusion constant or dispersion is generally defined as
\begin{equation}\label{D}
D=\frac{1}{2} \lim\limits_{t\rightarrow\infty}^{} \frac{d}{dt}\left[\langle x(t)^2\rangle -\langle x(t)\rangle^2\right].
\end{equation} 
From the definition (\ref{ukn}) and the summation of Eqs. (\ref{u1}) - (\ref{u3}) it follows that 
\begin{eqnarray}\label{dx2}
\frac{d}{dt}\langle x(t)^2\rangle=\frac{d}{dt}\sum\limits_{y=-\infty}^{+\infty}\sum\limits_{k=0}^{\infty}\sum\limits_{i=0}^{N-1} (yN+kN+i)^2 P_{yN+kN+i,y}(t)=
\nonumber \\
\frac{d}{dt}\sum\limits_{k=0}^{\infty}\sum\limits_{i=0}^{N-1}U_{kN+i}(t)=\sum\limits_{k=0}^{\infty}\sum\limits_{i=0}^{N-1}\frac{dU_{kN+i}(t)}{dt}=2S_0(t)+2-R_0(t),
\end{eqnarray} 
which implies that we do not need to actually solve the system (\ref{u1}) - (\ref{u3}) to get the diffusion coefficient. In the derivation of Eq. (\ref{dx2}) we also used the fact that the probabilities $R_{kN+i}(t)$ are normalized. In the stationary-state limit we have $S_0(t)\rightarrow a_0 t+T_0$, where $a_0 =R_0^2$ [from Eq. (\ref{akn})] and $T_0=B-\Delta$ [from Eq. (\ref{tt0})]. Also, $R_0(t)\rightarrow R_0$ at steady-state conditions. Therefore,
\begin{equation}\label{dx2dt}
\frac{d}{dt}\langle x(t)^2\rangle=2R_0^2 t+2B-2\Delta+2-R_0.
\end{equation}
Taking into account that in the large-time limit $\frac{d}{dt}\langle x(t)\rangle=\sum\limits_{k=0}^{\infty}\sum\limits_{i=0}^{N-1}a_{kN+i}=R_0$ [see Eq. (\ref{vv})] and $S_{kN+i}(t)\rightarrow a_{kN+i}t+T_{kN+i}$ and using the expression for $\langle x(t)\rangle$ [Eq. (\ref{avx})], we obtain 
\begin{eqnarray}\label{dxdt2}
\frac{d}{dt}\langle x(t)\rangle^2 = 2\langle x(t)\rangle\frac{d}{dt}\langle x(t)\rangle = 2R_0 \langle x(t)\rangle=2R_0 \sum\limits_{k=0}^{\infty}\sum\limits_{i=0}^{N-1}S_{kN+i}(t\rightarrow\infty)= \nonumber \\
2R_0 \sum\limits_{k=0}^{\infty}\sum\limits_{i=0}^{N-1}\left\{a_{kN+i}t+T_{kN+i}\right\}=2R_0^2 t+2R_0 \sum\limits_{k=0}^{\infty}\sum\limits_{i=0}^{N-1}T_{kN+i}.
\end{eqnarray} 
Substituting the results (\ref{dx2dt}) and (\ref{dxdt2}) into Eq. (\ref{D}) yields the diffusion constant,
\begin{equation}\label{DD}
D=\frac{1}{2} \left[2B-2\Delta+2-R_0-2R_0\sum\limits_{k=0}^{\infty}\sum\limits_{i=0}^{N-1}T_{kN+i}\right].
\end{equation}  
Note that in this expression all time-dependent terms disappear, as expected, confirming our initial assumptions.

To obtain the final expression for the dispersion in Eq. (\ref{DD}) we have to compute the term with $\sum\limits_{k=0}^{\infty}\sum\limits_{i=0}^{N-1}T_{kN+i}$. It can be shown from Eq. (\ref{tt0}) that
\begin{eqnarray}\label{sumT}
\sum\limits_{k=0}^{\infty}\sum\limits_{i=0}^{N-1}T_{kN+i}=\sum\limits_{k=0}^{\infty}\sum\limits_{i=0}^{N-1}x^k \left \{(Ak+B)\left[1-\frac{i}{N}(1-x)\right]-\frac{i(i+1)}{2}\Gamma \right.  \nonumber \\
\left. -\frac{i(i^2-1)}{6}\frac{(1-x)}{N}R_0^2 -\Delta \right \},
\end{eqnarray}  
Utilizing the fact that 
\begin{equation}\label{}
\sum\limits_{k=0}^{\infty}\sum\limits_{i=0}^{N-1}Bx^k \left[1-\frac{i}{N}(1-x)\right]=\frac{B}{2(1-x)}[N+1+(N-1)x]=\frac{B}{R_0},
\end{equation}
[as follows from Eqs. (\ref{rr0}) and (\ref{x})] and
\begin{equation}\label{}
\sum\limits_{k=0}^{\infty}\sum\limits_{i=0}^{N-1}Akx^k \left[1-\frac{i}{N}(1-x)\right]=\frac{Ax}{2(1-x)^2}[N+1+(N-1)x],
\end{equation}
yields  Eq. (\ref{sumT}) taking the form,
\begin{eqnarray}\label{TT}
\sum\limits_{k=0}^{\infty}\sum\limits_{i=0}^{N-1}T_{kN+i}=\frac{Ax}{2(1-x)^2}[N+1+(N-1)x]+\frac{B}{R_0}- \frac{N(N^2-1)}{6(1-x)}\Gamma \nonumber \\
- \frac{(N-2)(N^2-1)}{24}R_0^2-\frac{N}{1-x}\Delta,
\end{eqnarray}  
where we also took into account that $0\le x<1$ (for $p\neq 0$) and used the following known results, $\sum\limits_{i=0}^{N-1}i(i+1)=\frac{N(N^2-1)}{3}$ and  $\sum\limits_{i=0}^{N-1}i(i^2-1)=\frac{N(N-2)(N^2-1)}{4}$.

 Finally, substituting Eq. (\ref{TT}) into Eq. (\ref{DD}) leads to the following result for the diffusion coefficient $D$,
\begin{eqnarray} \label{finalD}
D=\frac{1}{2}\left[-2\Delta+2-R_0-2R_0 \left \{\frac{Ax}{2(1-x)^2}[N+1+(N-1)x]-\frac{N(N^2-1)}{6(1-x)}\Gamma \right. \right. \nonumber \\
\left. \left. -\frac{(N-2)(N^2-1)}{24}R_0^2-\frac{N}{1-x}\Delta \right \} \right].
\end{eqnarray}  
In this expression the undetermined parameter $B$ cancels out and does not affect the result for $D$, as was mentioned above. In the Eq. (\ref{finalD}) parameters $A$, $\Delta$, $\Gamma$, $R_0$, $x$ are given explicitly by  Eqs. (\ref{A}), (\ref{delta}), (\ref{gamma}), (\ref{rr0}), and (\ref{x}), correspondingly; the parameter $\alpha$ in the expressions for $A$ and $\Delta$ is given by Eq. (\ref{alpha}). Since all these parameters are functions of only the concentration of bridges $c$ and the burning probability $p$, Eq. (\ref{finalD}) provides the exact and explicit expression for $D(c,p)$.

\section{Illustrative Examples}

To illustrate our theoretical approach we compute dynamic properties of single motor protein in BBM for several sets of parameters. First, let us consider a simple case of $p=1$, i.e., when any crossing of the bridge burns it. It can be easily shown that
\begin{equation}
x=0, \quad R_0=\frac{2}{N+1}, \quad  \Gamma=\frac{4}{N(N+1)^2}, \quad \alpha=\frac{2}{3}\frac{(N-1)}{(N+1)}.
\end{equation}
Parameters $A$ and $\Delta$ have removable singularity at $p=1$ (and therefore so does $D$). Thus we need to compute $\lim\limits_{p\rightarrow 1}^{}A$ and $\lim\limits_{p\rightarrow 1}^{}\Delta$. To do so  we introduce a parameter $\varepsilon \ll 1$ so that $p=1-\varepsilon$. Then $x=\frac{\varepsilon}{N+1}+O(\varepsilon^2)$ as $p\rightarrow 1$. Using  this relation  we obtain
\begin{equation}
\lim\limits_{p\rightarrow 1}^{}A=\frac{2}{3}\frac{N(2N+1)}{(N+1)^2}, \quad \lim\limits_{p\rightarrow 1}^{}\Delta=-\frac{2}{3}\frac{(N+2)}{(N+1)}.
\end{equation}
Then, according to Eq. (\ref{finalD}) and Eq. (\ref{vv}), the dynamic properties of motor proteins are given by
\begin{equation} \label{Vp1}
V=\frac{2}{N+1}=\frac{2c}{c+1},
\end{equation}
and
\begin{equation}\label{Dp1}
D=\frac{2}{3}\frac{(N^2+N+1)}{(N+1)^2}=\frac{2}{3}\frac{(c^2+c+1)}{(c+1)^2}.
\end{equation}
These expressions  coincide with those obtained in \cite{morozov07}  where the velocity and diffusion constant  were computed for periodic bridge distribution with $p=1$ from the original Derrida's method, confirming the validity of our theoretical method.

Another interesting example is $c=1$, when every bond in the lattice is a potential bridge that can be burned. In this case all auxiliary functions can be easily obtained,
\begin{equation}
x=1-\sqrt{p}, \quad R_0=\sqrt{p}, \quad \Gamma=p, \quad \alpha=0, \quad A=\frac{\sqrt{p}}{2}, \quad \Delta=-\frac{\sqrt{p}+1}{2}. 
\end{equation}
Substituting these results into Eqs. (\ref{vv}) and (\ref{finalD}), we obtain for  the dynamic properties the following expressions,
\begin{equation}\label{Dc1}
V=\sqrt{p}, \quad D=\frac{1}{2}.
\end{equation}
It is interesting to note that in this case the mean velocity depends on the burning probability, while the dispersion is independent of $p$. When $p=0$ the diffusion constant is equal to one because it corresponds to unbiased motion of the random walker. However, any nonzero probability of burning changes the dynamic behavior at large times significantly by lowering fluctuations. Thus for any $p>0$ the particle is always in the biased diffusion regime in the stationary-state limit.

Generally, in the case of very low burning probability, i.e., when $p\ll c$,   we have the following expressions for the auxiliary functions (to the leading order in $p/c$),
\begin{equation} 
x \simeq 1-\sqrt{p/c}, \quad R_0 \simeq \sqrt{pc}, \quad \Gamma \simeq pc, \quad \alpha \simeq \frac{(1-c)p}{2c}, \quad A \simeq \frac{1}{2}\sqrt{p/c}, \quad \Delta \simeq -\frac{1}{2}. 
\end{equation}
Then the dynamic properties are equal to
\begin{equation}
V \simeq \sqrt{pc}, \quad D \simeq \frac{1}{2}.
\end{equation}

In the opposite limiting case, when $c\ll p$, our calculations to the leading order in $c/p$ produce,
\begin{equation} 
x \simeq c(\frac{1}{p}-1), \quad R_0 \simeq 2c, \quad \Gamma \simeq \frac{4c^3}{p},\quad  \alpha \simeq \frac{2}{3}, \quad  A \simeq \frac{4}{3}, \quad \Delta \simeq -\frac{2}{3}. 
\end{equation}
The corresponding expressions for the mean velocity and dispersion are given by
\begin{equation}\label{}
V \simeq 2c, \quad D \simeq \frac{2}{3}.
\end{equation}
The value for the dispersion is rather unexpected since in the case of no bridges ($c=0$) we have $D=1$, as expected for the unbiased diffusion of the random walker.

\section{Discussions}

To test our theoretical approach,  Monte Carlo computer simulations of dynamic properties of motor proteins in the burnt-bridge model have been developed. We performed computer simulations  by using  methods discussed in detail in Ref. \cite{morozov07}. Since the velocity of the motor protein in BBM  has already been determined in analytical calculations and Monte Carlo simulations, we concentrated our efforts on computations of dispersion $D(c,p)$. Diffusion constants or dispersions have been calculated in simulations via the formula $D=(\langle x^2\rangle -\langle x \rangle^2)/2t$.  Since the results of computer simulations for dispersion indicate large fluctuations, especially for very low $c$ and $p$, typically $10^5$-$10^6$ simulation runs have been averaged over to decrease stochastic noise.  

The results of Monte Carlo computer simulations for dispersions with different parameters are presented in Fig. 2. Comparison of numerical computations with analytical predictions indicates a very good agreement for all conditions, and it  validates our theoretical method. As shown in Fig. 2a,   $D(c)$ is a monotonically decreasing function for any fixed value of $p \neq 0$, and  it decreases more slowly for larger values of burning probability $p$. This behavior is expected since the burning of the bridges limits the mobility of the hopping particle. At very low concentration of bridges ($c \ll p$) dispersion for all values of $p \ne 0$ approached a limiting value of 2/3, in agreement with analytical predictions. However, this limiting value is less than $D=1$ for the case of the diffusion on the lattice without bridges.  Note also large fluctuations in $D(c)$ obtained from Monte Carlo simulations for $c \ll 1$ due to the fact that the system, probably, has not reached a stationary state yet.

Dispersion as a function of the burning probability shows a more complex dependence, as illustrated in Fig. 2b. For $p=0$ the dispersion is equal to one, as expected, for any $c$. The increase in $p$ first significantly lowers the dispersion and then $D(p)$ starts to increase. This non-monotonic behavior is rather unexpected because one can naively argue that the increase in the burning probability should limit fluctuations of the particle, leading to the decrease in the diffusion constant. One can understand the dynamics of the single particle in BBM in the following way. Irreversible burning of the bridges limits the motion of the particle in one direction, but it allows the particle to move further to the right until it crosses another bridge that will burn. It leads to larger effective hopping rates, resulting in the larger dispersions. Note that for $c=1$ the dispersions is constant and equal to 1/2 for any $p >0$, in agreement with theoretical predictions. In addition, $D(p) \rightarrow 1/2$ for very low probabilities of burning ($p \ll c$) and $c \ne 0$.

\begin{figure}[tbp]
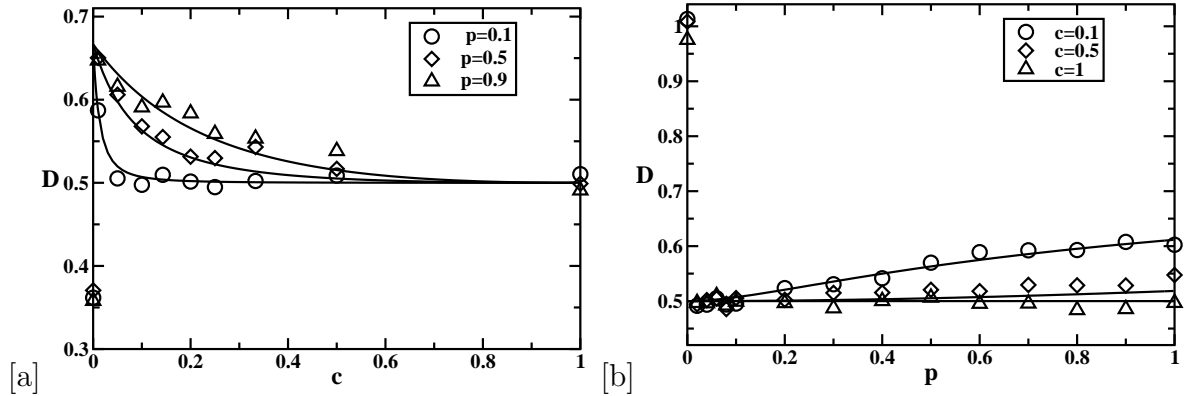
\label{fig2}
\centering
[a]\includegraphics[scale=0.3,clip=true]{Fig2a.bb2.eps}
[b]\includegraphics[scale=0.3,clip=true]{Fig2b.bb2.eps}
\caption{Diffusion constants of the particle in the burnt-bridge model with periodically distributed bridges: a) as a function of concentration of bridges; and b) as a function of burning probability. Solid lines are results of analytical calculations. Symbols are from Monte Carlo computer simulations.}
\end{figure}

The presented theoretical analysis provides a full description of dynamics of motor protein in BBM. Most observed trends in the dynamic properties can be well understood, however there are several surprising observations. First, in the limit of very low concentration of bridges ($c \ll p$) our analytical calculations and computer simulations  yield $D(p)=2/3$ which is not equal to  $D=1$ expected for the case of $c=0$. This jump in the dispersion is a signature of the dynamic phase transition \cite{morozov07}, that separates unbiased diffusion regime ($c=0$) from biased diffusion behavior ($c>0$). Another interesting observation is the fact that $D=1/2$ for $c=1$ for {\it any} value of the burning probability $p$. This can be understood by analyzing the trajectories of the particle. After the bridge is burned the random walker spends most of its time near the burning position that behaves like a hard wall and it rarely diffuses forward. This lowers significantly fluctuations in the system, and the dynamics of the particle become independent of the burning probability $p$.

\section{Summary and conclusions}

A comprehensive theoretical approach to describe the dynamics of motor protein particles in one-dimensional burnt-bridge model with periodically distributed bridges  is presented. Our theoretical method allows one to calculate exactly all dynamic properties of the system for any set of parameters, and specific calculations are given for the mean velocity and dispersion. It is found that dispersion is monotonically decreasing function of the concentration of weak links $c$, while the dependence of $D$ on the burning probability $p$ is more complex. Trends in dispersions are discussed using simple physical arguments. Theoretical analysis also indicates that there is a gap in the dispersions in the limit of $c \rightarrow 0$, that can be associated with a dynamic phase transition between biased and unbiased diffusion regimes. In addition, our calculations show that for any concentration of bridges the fluctuations of the particle are lowered. This is due to the fact that the particle spends most of the time near the last burned bridge that acts as a hard wall. Our theoretical findings are confirmed by extensive Monte Carlo simulations.

In this paper only periodic distribution of weak links have been considered. This situation is realistic for collagenases moving along collagen fibrils \cite{saffarian04}, although different distributions of weak links are also generally possible.  It will be interesting to investigate BBM with random distribution of bridges, for which limited theoretical results are available \cite{antal05}.  It will also be  important to generalize our theoretical approach for studying the dynamics of dimers moving on the parallel lattices in BBM. There are theoretical predictions that dimers are more effective molecular motors than monomers \cite{saffarian06}. In addition, our theoretical method provides an explicit analysis of dynamic properties of random walkers in quasi-periodic lattice, and this results can be applied for investigation of  motor protein dynamics in other systems \cite{AR}.

\ack
 
The support  from the Welch Foundation (under Grant No. C-1559), and from the US National Science Foundation through the grant CHE-0237105 is gratefully acknowledged. A.B.K. and M.N.A. are grateful to Yulia Ivanova for support and stimulating discussions.

\end{document}